\documentclass[prb,preprint]{revtex4-1}

\usepackage{amsmath}
\usepackage{graphicx}
\usepackage{natbib}
\usepackage{lipsum}

\def\dbar{{\mathchar'26\mkern-12mu d}}

\begin{document}

\title{Thermodynamics. Using Affinities to define reversible processes.}

\author{Hern\'an A. Ritacco}
\email{hernan.ritacco@uns.edu.ar}
\affiliation{Instituto de F\'isica del Sur (IFISUR), Consejo Nacional de Investigaciones Cient\'ificas y T\'ecnicas (CONICET)  and Departamento de F\'isica, Universidad Nacional del Sur (UNS). Av LN Alem 1253 (8000) Bah\'ia Blanca, Argentina.}
\date{\today}

\begin{abstract}
In this article a definition of reversible processes in terms of differences in intensive Thermodynamics properties (Affinities) is proposed. This definition makes it possible to both define reversible processes before introducing the concept of entropy and avoid the circularity problem that follows from the Clausius definition of entropy changes. The convenience of this new definition compared to those commonly found in textbooks is demonstrated with examples. 
\end{abstract}

\maketitle

\section{Introduction}

Thermodynamics is a subtle subject involving abstract concepts which are often difficult for students to grasp \cite{Granville1985, Smith2015, Langbeheim2013, Christensen2009a, Jasien2002, Cochran2006, Loverude2002, Bucy2006}. Among these concepts we find that of reversibility, which is central in Thermostatics (equilibrium Thermodynamics). In his book on the subject Duhem \cite{duhem1903thermodynamics} wrote ``In order to state this principle (of Thermodynamics) it is necessary for us to become acquainted with one of the most delicate principles in all Thermodynamics, that of reversible transformation". However important, reversibility is, in general, an ill-defined concept in textbooks. \\

\begin{figure}[tbh]
\centerline{\resizebox{0.45\linewidth}{!}
{\includegraphics{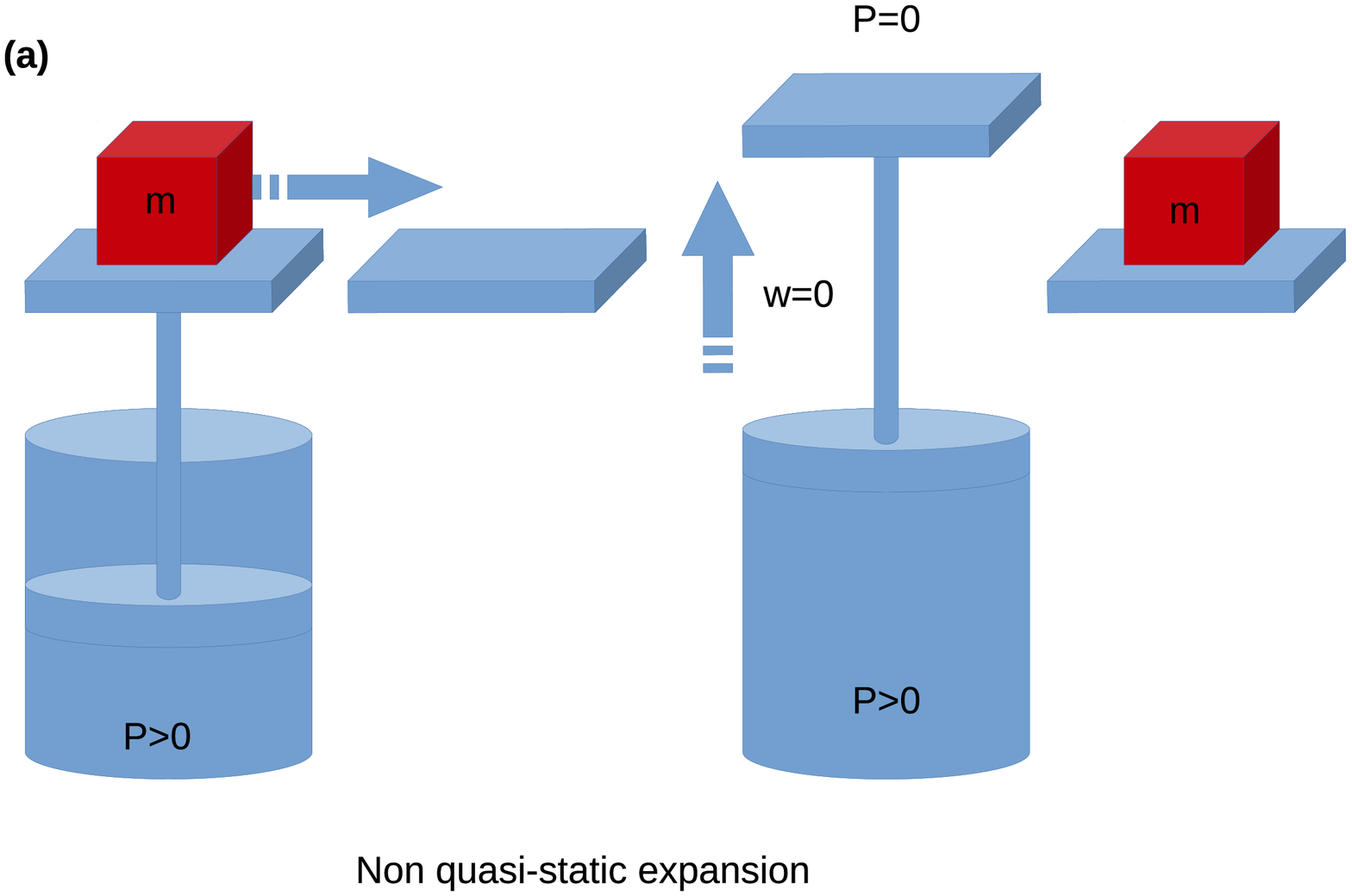}} \hspace{0.45cm}
\resizebox{0.5\textwidth}{!} {\includegraphics{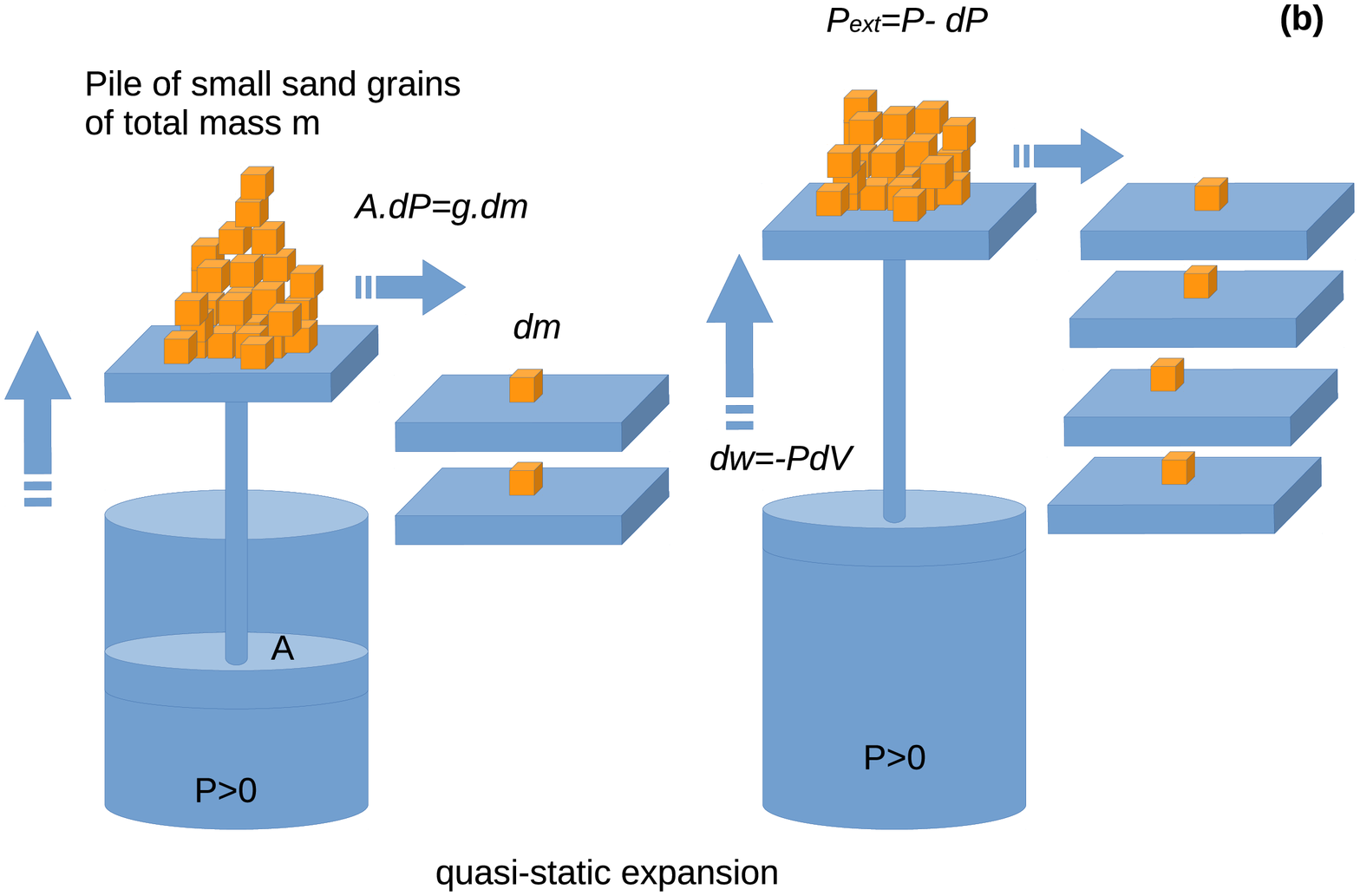}}}
\caption{The typical way of explaining reversible processes. (a) A piston filled with gas expands against zero pressure when the mass m is removed: irreversible non-quasi-static expansion. (b) The same piston expands against a pressure $p=p_i - dp$ in a reversible quasi-static process. In this case the mass m is replaced by a pile of sand. Each grain of sand is removed one by one to mimic the quasi-static path.}
\label{sand}
\end{figure}

Closely related with reversibility is the idea of quasistatic paths. Along these paths, the processes involved in the evolution of the system from the initial to the final state, advance infinitely slowly. It is common to introduce and explain reversible processes to students using the example shown in figure \ref{sand}, \cite{Meany1978, VanNess1983}. In the case of an expansion process driven by an infinitesimal difference in pressure, $dP$, as shown in Figure \ref{sand}b, we argue that, because the  process is quasistatic, the external pressure can be replaced by the systems pressure, and we could change the direction of the process just by changing the sign of $dP$. Thus the process is reversible. This direct association between a quasistatic path and a reversible process could induce us to think that any quasistatic process is reversible. Most introductory Thermodynamics, Physics and Physical Chemistry textbooks do not make a clear distinction between reversible and quasistatic paths. It has to be stressed here that a quasistatic locus is not necessarily reversible \cite{Thomsen1960a,Thomsen1960,Samiullah2007}. 
Examples as the one mentioned above lead to a commonly found definition for reversible transformations: ``...a change that can be reversed by an infinitesimal modification of a variable'', or its equivalent  ``...a process that can be reversed by means of only small changes in their environment...''  \cite{Atkins2010, Walker2013,Fermi1956,Smith1996}. These definitions do not specify what kind of \textit{variable} we are talking about nor what we mean by a small change in the environment. Similarly, what is meant with the word \textit{reversed} is not clear. These definitions are frequently advanced in the context of ``reversible expansion work '' calculations, well before introducing entropy and the second law of Thermodynamics. Below, I will illustrate with examples that this kind of definitions may conduct to confusing results when applied to certain problems.

The problem of the vague and misleading definitions for reversible processes is not new and has been addressed before. Rechel, for instance analyzed the problem in an article published in Journal of Chemical Education in 1947 \cite{Rechel1947}. In his paper, Rechel discusses and criticizes the definitions for reversible transformations from different books and concludes that, in order to avoid inconsistencies and contradictions with the fundamental principles of mechanics, reversible processes must be defined in terms of a limiting value of a mathematical function. However, Rechel does not clearly identify the mathematical function to be used, instead he utilizes the typical example of expansion of perfect gases and calculates the limiting value for the reversible expansion work. 
The function that must be used is of course the entropy (S). Landau-Lifshitz\cite{Landau1969}, Zemansky\cite{Zemansky1996} and Callen \cite{Callen1985} use this approach when defining reversibility. Callen defines ``...a quasistatic locus as a dense succession of equilibrium states...'' and then ``...the limiting case of a quasistatic process in which the increase of entropy becomes vanishingly small is called reversible process...''. In a similar approach, Thomsen and Bears \cite{Thomsen1996} propose a mathematical definition of a reversible path formulated in terms of ``a zero-entropy-production limit''. This way of defining a reversible path implies that both the second law of Thermodynamics and the concept of entropy have to be explained to students before introducing the concept of reversibility. It should be emphasized here that a formally correct definition for a reversible process has to be made in terms of zero entropy production of the process \cite{Samiullah2007}. A reversible process is one for which the total entropy change (systems + surroundings) is zero, $\Delta S = 0$. If the total entropy change is not zero there is no process capable of returning the system and its sourroundings (system+sourrondings form an isolated composed system) to their initial states. That is what the word \textit{reversed} really means when used in the definitions of Thermodynamics reversibility. Then, when discussing reversible processes, the second law cannot be really avoided although it was proposed differently\cite{Gislason2002}. 
However, this leads us to a deadlock because entropy is generally introduced in first courses by the Clausius definition: $dS=\frac{\dbar q_{rev}}{T}$, being $\dbar q$ the heat exchange and T the temperature at which the exchange takes place. The subscript ``rev'' means that the process of heat transfer must be calculated in a reversible path in order to obtain the entropy change in that process. Then, the definition of reversibility in terms of vanishing entropy production is in some way useless because we have to identify a reversible path in order to calculate the entropy change but, because the definition states that a reversible process is one where $\Delta S = 0$, we first have to calculate the total entropy change to ensure that the path we have chosen to calculate the entropy is reversible. This is no doubt very confusing and not only for students. The origin of this is historical: Carnot introduced the concept of \textit{reversible work, heat transfer} and \textit{reversible machines} before Clausius advanced the concept and definition of entropy producing this \textit{circularity problem}.

Because the concept of reversibility is central and has to be introduced at early stages in the study of Thermodynamics, the question is how we could give an operational definition for reversible processes before introducing the Second Law of Thermodynamics in order to avoid the circularity problem described above. This can be done by identifying the driving forces for the process in question. Before developing this idea let me present first some preliminary concepts.

\section{Preliminary concepts.}

\subsection{Extensive and Intensive variables}
An extensive parameter is one which depends on the system size. Let me illustrate this with an example. If we add to one litre of water at a temperature of 298 K another litre of water at the same temperature, the total final volume, V will be 2 litres. Then V is an extensive variable that depends on the amount of matter. On the other hand, the temperature of the final system of two litres of water is not 298 K + 298 K = 596 K, it is just 298 K. Temperature does not depend on the system size, it is an intensive parameter. Intensive parameters are in general relations between two extensive ones, for example, because density ($\rho$) is mass per volume, pressure (\textit{P}) is force per unit area and surface tension($\gamma$) is energy per unit area, then $\rho$, \textit{P} and $\gamma$ are three intensive parameters. 

\subsection{The unique problem in Thermostatics}
\begin{figure}[tbh]
\includegraphics[scale=0.35]{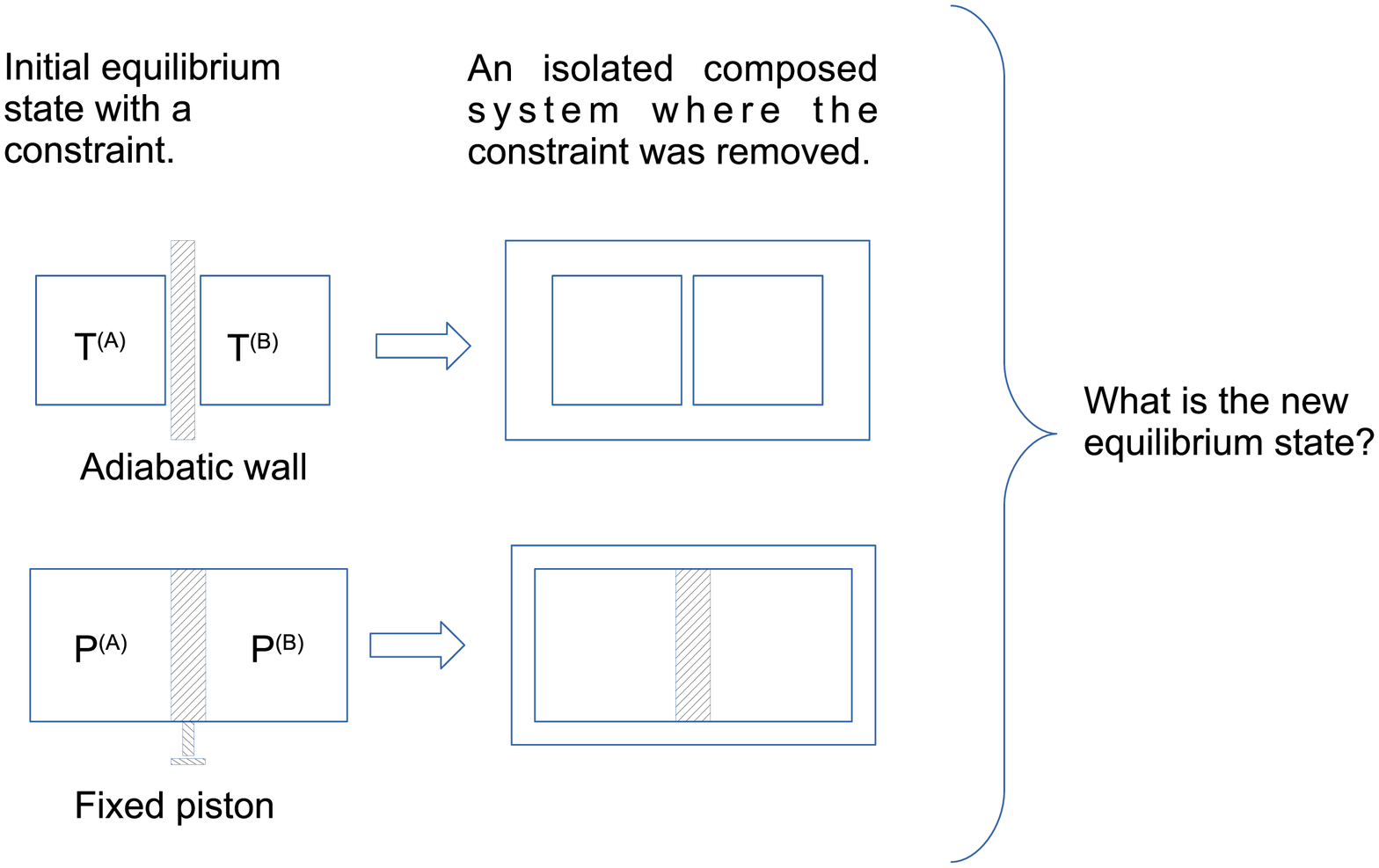}
\caption{(a) Thermal contact; (b) moving pistons.}
\label{thermoprobleme}
\end{figure}

The single problem of equilibrium thermodynamics is the determination of the equilibrium state that eventually results after the removal of internal constraints in a closed, composite system \citep{Callen1985}. All problems in Thermostatics can be reduced to this. Let me illustrate how with examples. Suppose we want to determine the final temperature of two bodies when they are put in thermal contact in an adiabatic process (figure \ref{thermoprobleme}a). We can think that initially the two bodies were in equilibrium separated by an internal constraint, an adiabatic wall, forming a composite system (composed by the two bodies in an isolated closed system). Then, the restriction is eliminated permitting the flux of energy as heat between the two bodies and forming a composite isolated system as shown in figure \ref{thermoprobleme}a. The problem reduces now to obtaining the new equilibrium state of our composite isolated system once the internal constraint is removed. Consider now another example: two gaseous systems separated by a diathermal piston fixed in place by a screw. We desire to obtain the equilibrium position of the piston when the screw is removed if the whole process is adiabatic. In this case we form a new composite isolated system as shown in figure \ref{thermoprobleme}b. Again the problem reduces to finding the new equilibrium condition of this isolated closed system. If the process is not adiabatic we just include a third subsystem, the \textit{surroundings}, to give place to an isolated composite system (with three subsystems) and proceed in the same way. The solution for the unique problem of Thermostatic is obtained by searching for the maximum of a certain function that describes the composite isolated system, once the internal constraints are removed. This function is the total entropy, S.    

\subsection{The formalism in thermostatics.}
 
By combining the Clausius definition for entropy change, $dS= \dbar q_{rev}/T$  with the first law of thermodynamics for a closed system, $dU=\dbar q + \dbar w$ (being U the Internal energy and w the work), and assuming, for now, that only pressure-volume (PV) work (quasi-static) can be done, $\dbar w = -P dV$ it follows that,

\begin{equation}
dU= T dS - PdV
\label{eq1}
\end{equation}

Written in this way, U can be viewed as some function of the extensive independent variables S and V, $U(S,V)$; then its differential will be

\begin{equation}
dU=\frac{\partial U}{\partial S} dS + \frac{\partial U}{\partial V}dV
\label{eq2}
\end{equation}

By comparing eqs. \ref{eq1} and \ref{eq2}, we can define T as the partial derivative of U(S,V) with respect to the entropy: $T=\partial U/\partial S$. From this, it is now clear that T is an intensive variable because U and S are extensive. In the same way we can define $P=-\partial U/\partial V$. Because the partial derivatives of any function are also functions of the same variables, then T(S,V) and P(S,V) give the relations between T and P with the independent variables S and V. These functions are called equations of state \cite{Callen1985}. It is neither common nor convenient to have S as an independent variable. Let us then rewrite eq. \ref{eq1} as

\begin{equation}
dS = \frac{1}{T}dU + \frac{P}{T} dV
\label{eq3}
\end{equation}

Now, we can say that S, in eq. \ref{eq3}, is a function of the independent variables U and V, $S(U,V)$. 
When the differential form is written as in eq. \ref{eq1}, we say that it is in the Energy representation, meaning that U is the dependent variable; when written as eq. \ref{eq3}, we say that it is in the Entropic representation, meaning that S is now the dependent variable. The function $S(U,V)$ is the \textit{Fundamental Relation} of our thermodynamic system. We can identify the partial derivatives in eq. \ref{eq3} as $\frac{\partial S}{\partial U} = \frac{1}{T}$ and $\frac{\partial S}{\partial V} = \frac{P}{T}$. These partial derivatives are also functions of the same independent variables as S(U,V).
Because the second principle of thermodynamics is expressed in terms of the maximum of the entropy, we assume (postulate) that for any thermodynamic system there is a fundamental relation $S(U,X_i)$, the entropy, which is a function of all extensive parameters, $X_i$, necessary to describe it correctly. That is, we postulate the existence of the integral of eq. \ref{eq3} for each thermodynamic system, and then we generalize the treatment to all systems by including in $S(U,Xi)$ all the relevant extensive parameters needed to describe those systems completely. In reference \onlinecite{Ritacco2014} an example of the application of this formalism, in which the authors work out the fundamental equation of a non-ideal rubber band, is presented. Since entropy is additive, the total entropy of a composite system is the sum of the entropy of each subsystem, $S=\sum_j S^{(j)}(U^{(j)},X_i^{(j)})$, at equilibrium the parameters $U^{(j)}, \; X_i^{(j)}$ adopt values that maximize the total entropy, $S= \sum_j (S^{(j)})$.

\subsection{Equilibrium, Driving forces for change and Affinities.}
\begin{figure}[tbh]
\includegraphics[scale=0.35]{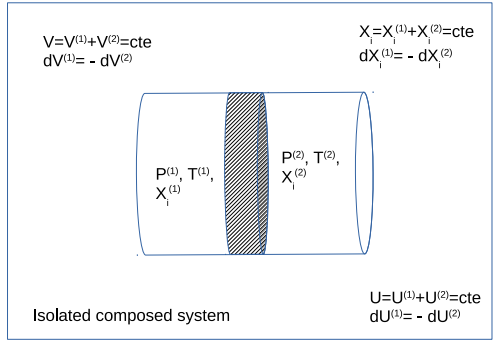}
\caption{Isolated composite system}
\label{composedsys}
\end{figure}

The fundamental equation S(U, $X_i$) contains all the thermodynamics information of the system. Suppose we have an isolated composite system formed by two subsystems separated by a completely restrictive wall as shown in figure \ref{composedsys}. With \textit{completely restrictive wall} we mean that the wall does not allow any redistribution of extensive variables between subsystems: i.e the wall is adiabatic and energy cannot be transferred as heat between the subsystems, it is rigid and the volume of subsystems cannot change, it is non permeable to any substance and then the mole numbers in each subsystem remain constant, etc. A process initiates when one, some or all of these restrictions are removed, and our task is to determine the new equilibrium state for the composite system that eventually results. The second principle states that the total entropy, $S(U^{(1)}, U^{(2)},X_i^{(1)}, X_i^{(2)})$ should be a maximum at equilibrium ($dS=0$, $d^2S<0$),

\begin{widetext}

\begin{equation}
dS= \frac{1}{T^{(1)}}dU^{(1)} + \sum_i \left( \frac{\partial S^{(1)}}{\partial X_i^{(1)}} dX_i^{(1)}\right) + \frac{1}{T^{(2)}}dU^{(2)}+ \sum_i \left( \frac{\partial S^{(2)}}{\partial X_i^{(2)}} dX_i^{(2)}\right) = 0 \;\;\; (at \, equilibrium.)
\label{eqeq}
\end{equation}

\end{widetext}

Let me consider just two variables, U and V, in eq. \ref{eqeq} and the process of figure \ref{composedsys} in order to make the presentation of ideas more explicit and clear. In that case the equilibrium condition for the isolated composite system reduces to,

\begin{equation}
dS = \frac{1}{T^{(1)}}dU^{(1)} + \frac{P^{(1)}}{T^{(1)}}dV^{(1)} + \frac{1}{T^{(2)}}dU^{(2)} + \frac{P^{(2)}}{T^{(2)}}dV^{(2)}=0
\end{equation}

Because both the total internal energy (isolated composite system) and the total volume are constant, we have the following closure relations, $dU^{(1)}=-dU^{(2)}$; $dV^{(1)}=-dV^{(2)}$ then,

\begin{equation}
dS = \left[ \frac{1}{T^{(1)}}- \frac{1}{T^{(2)}} \right] dU^{(1)} + \left[ \frac{P^{(1)}}{T^{(1)}}-\frac{P^{(2)}}{T^{(2)}} \right] dV^{(1)} = 0
\label{eqcond}
\end{equation} 

since at equilibrium $dS=0$ independently of $dU,\; dV$ the brackets in eq. \ref{eqcond} must be zero and thus, the equilibrium conditions are $T^{(1)}=T^{(2)}$ and $P^{(1)}=P^{(2)}$, as expected.\\
Now, in a quasistatic stepped process we can replace $dS$ by $\Delta S$ in each step of the path from the initial to the final state of the composite system,

\begin{eqnarray}
\Delta S_{step} = \left[ \frac{1}{T^{(1)}}- \frac{1}{T^{(2)}} \right] \Delta U^{(1)} + \left[ \frac{P^{(1)}}{T^{(1)}}-\frac{P^{(2)}}{T^{(2)}} \right] \Delta V^{(1)} \nonumber \\ 
\label{step}
\end{eqnarray} 

where $\Delta S_{step}$ has to approach zero for all steps if the process is to be reversible ($\Delta S = 0$), independently of $\Delta U$ and $\Delta V$. As a consequence the brackets have to tend to zero in each step for a reversible process. The terms inside the brackets are the generalized force, called affinities, that drives any irreversible process from a non-equilibrium to an equilibrium state. In simple systems, they are generally expressed as a difference in an intensive variable among subsystems. The variables involved in this case are T and P, but the analysis can be easily generalized (eq. \ref{eqeq}) to other intensive variables and processes. Summarizing, a process driven by an arbitrarily small affinity guarantees that the entropy production of the process tends to zero, such a process approaches a thermodynamic reversible process.
 
\section{A definition of a Reversible path in terms of Affinities.}
We could then define a reversible process as follows: \textit{a process approaches a reversible transformation when it is driven by a vanishing affinity}. For simple systems this means that the process is driven by an infinitesimal difference in an intensive variable such as T, P, $\mu$ (chemical potential, $\mu= \partial U/ \partial N$), etc. 

\section{Application Examples.}

I will illustrate the problems that could be encountered when using the common definitions found in textbooks and show how the definition proposed above avoids them.

\subsection{Adiabatic Free expansion of an ideal gas.}
\begin{figure}[tbh]
\centering
\includegraphics[scale=0.30]{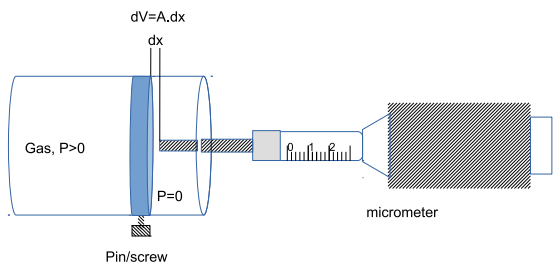}
\caption{The continuous free expansion of an ideal gas.}
\label{freeexp}
\end{figure}

The free expansion of a gas refers to an increase of its volume against zero pressure and is a typical example used when teaching Thermodynamics. Typically this is explained in terms of cylinders with moving frictionless pistons and the problem consists of calculating the change in the internal energy and entropy of the expansion process. The typical analysis is as follows. Because the external pressure is equal to zero, the work done, w (expansion work against the external pressure $P_{ext}$) is also zero, $\dbar w = - P_{ext} dV = 0 $. Then, from the first law of Thermodynamics for a closed system: $dU=\dbar q + \dbar w = \dbar q$, and because the process is adiabatic, $\dbar q = 0$. It follows that $dU=0$. For a perfect gas $dU=cNRdT$ (c is a positive constant, N the number of moles and R the ideal gas constant), it follows that $dT= 0$ and the temperature is constant during the expansion process.
For the calculation of the entropy change of the process we use the Clausius definition $dS= \dbar q_{rev}/ T$, which must be calculated for any reversible path from the initial to the final state. Then we need to think of a reversible path to carry out the sum $\sum_{n=\infty}( \dbar q_{rev}/T)$. Let us consider then the situation depicted on figure \ref{freeexp}. There we have a cylinder with our perfect gas inside with a moving (frictionless) piston in contact with an evacuated vessel, ($P_{ext} = 0$), the piston is initially fixed in place by a pin or screw which is then removed to let the system expand a $dV = A.dx$, where $dx$ is fixed by the micrometer and A is the piston area. After the first differential expansion dV, the pin is put back in place and the micrometer is turned in an additional dx to repeat the process until the final volume is reached. The distance dx can be as small as desired defining a continuous expansion such that $V=V_0+dV$ and we have a quasistatic locus. Now, we can think that we have defined a quasistatic and reversible path from initial to final state because a thermodynamic variable, the volume V, is changed in a dV. 

Here is the problem: If we try calculating the change in entropy of this process using the apparently reversible path of figure \ref{freeexp}, we immediately become confused because $P_{ext}=0$, and then $w = 0$ and $q = 0$ in each step of the expansion. Then, if the Clausius statement is correct, $S=\sum dS=\sum \dbar q_{rev}/T=0$ in an expansion! We  know that the change in entropy is in fact given by, $\Delta S=NR ln(V_f/V_i)$, being $V_f$ and $V_i$ the final and initial volumes, thus we have found an inconsistency.
Let us now apply the definition given in the previous section for a reversible path. If we use it in the case of figure \ref{freeexp} to determine if the chosen path from the initial to the final state is reversible or not, it is immediately clear that it is not. The process is driven by a finite difference in the intensive variable responsible for the change in volume, the pressure, the affinity $P_{int}-P_{ext} = P_{int}-0 \neq dP$, and the chosen path cannot be used to calculate the entropy change. The origin of the problem is the fact that the quasistatic path was defined in terms of the diferential change in an extensive variable, the volume, and this cannot be done (the driving forces for change are always differences on intensive variables). 
The definition in terms of affinities dictates how to correctly choose a reversible path without making use of the entropy concept and avoiding the circularity: In order to calculate the entropy change using the Clausius definition and a reversible path the affinity that drives the process has to approach zero: $P_{int}-P_{ext}= dP$, then we can replace the external pressure by the internal one given by the equation of state for the ideal gas: $P=NRT/V$. Because $dU=0$ then $\dbar q = -\dbar w = P.dV= (NRT/V) dV$ and $dS=\dbar q_{rev}/T= (NR/V)dV$ which integrated between $V_f$ and $V_i$ gives $\Delta S= NR ln(V_f/V_i)$.    

\subsection{Quasistatic transfer of heat.}

\begin{figure}[tbh]
\centering
\includegraphics[scale=0.35]{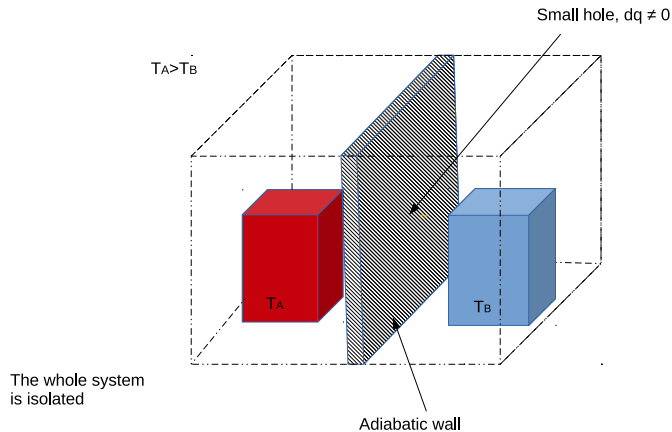}
\caption{Continuous transfer of heat}
\label{qquasi}
\end{figure}

Let us consider now another situation, depicted in figure \ref{qquasi}. We have two bodies, A and B at different temperatures, $T_A$ and $T_B$ , with $T_A > T_B$. The whole system is isolated and the bodies are separated by a perfect adiabatic wall, but this wall has a very small hole without thermal isolation that can be made as small as we want. This allows the flux of energy as heat at a rate that depends on the size of the hole and that can be as slow as we desire. In this way we have a quasistatic continuous transfer of energy as heat from A to B and we may think that we have defined a reversible locus to equilibrium since a variable, $U_a$ (and $U_B$, is changed in dU. Now if this is true and because the whole system is isolated, the infinitesimal change in the total entropy should be,

\begin{equation}
dS=\frac{\dbar q^A}{T_A}+\frac{\dbar q^B}{T_B}
\end{equation}

now, from the first law, $-\dbar q_A = \dbar q_B$ and because, until equilibrium is reached, $T_A> T_B$ always holds, then

\begin{eqnarray}
\Delta S &=& \int dS =\sum_{n=\infty} \frac{\dbar q_A}{T_A}+\frac{\dbar q_B}{T_B}=\sum_{n=\infty} -\frac{\dbar q_B}{T_A}+\frac{\dbar q_B}{T_B} = \nonumber \\
&=& \sum_{n=\infty} \left[ \frac{-1}{T_A}+\frac{1}{T_B} \right] \dbar q_B > 0
\end{eqnarray}

which means that the entropy increases even though the process as a whole is quasistatic in the energy (heat) transfer and adiabatic (the system composed by the two subsystems is isolated). Then, even though infinitely slow changes take place (we can make the hole as small as we want) the whole process is intrinsically irreversible ($dS > 0$).\\  
Again this problem originates because the definition of our quasistatic locus was done in terms of an extensive variable, the energy. If we use the definition in terms of affinities we find that the driving force for the energy exchange, the difference in temperatures, is $T_A-T_B \neq dT$ and then the process, even if quasistatic in energy transfer, cannot be reversible. Then we cannot calculate the entropy change of this process using the Clausius definition: $\sum \frac{\dbar q}{T}$ because the chosen locus is not reversible. Again the correct choice of path is dictated by the condition of a vanishing affinity, in this case $T_A-T_B=dT$.

\subsection{A case with friction forces.}

Let us see here how our definition applies in an example with frictional forces. Imagine the typical system of a gas inside a cylinder in a configuration similar to that of figure \ref{sand}b. The only difference is that we will now consider that sliding friction forces exist between the piston and the cylinder walls. In this case the process cannot be reversible \cite{Ferrari2010, Sherwood1984}, and our definition has to show this. In the calculation of the force responsible for the movement of the piston we need to consider the frictional force. For simplicity we will model it with the expression of dry friction: $F_f = \beta. f_N$; being $f_N$ the normal force and $\beta$ the friction coefficient. Now the net force for the movement of the cylinder is given by the driving force exerted on the piston by the pressure difference inside and outside it multiplied by the piston area, A, and the frictional force in such a way that: $A.(P_{out} - P_{in}) = A. \Delta P \geq \beta f_N$. Then the driving force, $\Delta P = P_{out}-P_{int}$, can approach the vanishing limit, $dP$, that from our definition guarantees reversibility, only if the normal, $f_N$, or the friction coefficient, $\beta$, approaches zero. That means that the process can be reversible only if there is no friction. If the frictional force has a finite value, $\Delta P$ cannot be arbitrarily small and it cannot tend to zero because, if the piston is going to move, the minimum value for $\Delta P$ should be at least equal to $\beta f_N/A$, and then the requisite of arbitrarily small driving force, a differential affinity, is not fulfilled and the process is irreversible. Note that in this case the process can be quasistatic by taking $\Delta P= \beta f_N/A$, but not reversible. For a very comprehensive treatment of this problem see reference \onlinecite{Besson2001}, here our goal is just to show how our definition for reversibility helps us see that a process with frictional forces cannot be reversible.


\section{Conclusion}

Because the driving forces for change are always differences between intensive variables, a process is reversible only when this differences are inffinitely small. The definition based on affinities helps avoid confusion when dealing with problems such as those presented here, and helps recognize if a process is reversible. If this is so then we can obtain the entropy change by means of the Clausius definition avoiding the circularity problem that emerges from it. The definition for reversible processes proposed in this article can be used in introductory thermodynamics courses without the need to demonstrate mathematically that it implies the zero entropy production limit, this can be left for more advanced courses or for when the concept of entropy is explained.\\
As a final comment let me state here that for more complex systems, affinities can involve a combination of different intensive variables of the subsystems depending on the closure conditions specific for each problem. However the requisite of a vanishing affinity still holds in order to have a reversible process.

\bigskip
{\em Acknowledgement:~~~} I thank Professors RG. Rubio and P. Schulz for very useful comments and critics on the manuscript. I thank Ayelen Prado for reading the manuscript and helping me to improve the English. This work was partially supported by grants PGI-UNS 24/F067 and PGIMayDS 24/MA09 of Universidad Nacional del Sur and PICT 2013 (D) Nro 2070 of Agencia Nacional de Promoci\'on Cient\'ifica y Tecnol\'ogica (ANPCyT). I thank NR and MFF for their continuous support.


\begin{thebibliography}{99}

\bibitem{Granville1985}{Granville, M. F., ``Student Misconceptions in Thermodynamics.'' J. Chem. Educ. \textbf{62} (10), 847--848 (1985).}

\bibitem{Smith2015} {Smith, T. I.; Christensen, W. M.; Mountcastle, D. B.; Thompson, J. R., ``Identifying Student Difficulties with Entropy, Heat Engines, and the Carnot Cycle.'' Phys. Rev. Spec. Top. - Phys. Educ. Res. \textbf{11} (2), 20114--20116 (2015).}

\bibitem{Langbeheim2013}{Langbeheim, E.; Safran, S. A.; Livne, S.; Yerushalmi, E., ``Evolution in Students’ Understanding of Thermal Physics with Increasing Complexity.'' Phys. Rev. Spec. Top. - Phys. Educ. Res. \textbf{9} (2), 020117 (2013).} 

\bibitem{Christensen2009a}{Christensen, W. M.; Meltzer, D. E.; Ogilvie, C. a., ``Student Ideas Regarding Entropy and the Second Law of Thermodynamics in an Introductory Physics Course.'' Am. J. Phys.  \textbf{77} (10), 907--917 (2009).}

\bibitem{Jasien2002} {Jasien, P. G.; Oberem, G. E., ``Understanding of Elementary Concepts in Heat and Temperature among College Students and K-12 Teachers.'' J. Chem. Educ. \textbf{79} (7), 889--895 (2002).}

\bibitem{Cochran2006} {Cochran, M. J.; Heron, P. R. L., ``Development and Assessment of Research-Based Tutorials on Heat Engines and the Second Law of Thermodynamics.'' Am. J. Phys. \textbf{74} (8), 734--741 (2006).}

\bibitem{Loverude2002} {Loverude, M. E.; Kautz, C. H.; Heron, P. R. L., ``Student Understanding of the First Law of Thermodynamics: Relating Work to the Adiabatic Compression of an Ideal Gas.'' Am. J. Phys. \textbf{70} (2), 137–148 (2002).}

\bibitem{Bucy2006} {Bucy, B. R., ``What Is Entropy? Advanced Undergraduate Performance Comparing Ideal Gas Processes.'' in \textit{AIP Conference Proceedings} 2006; Vol. 818, pp 81--84.}

\bibitem{duhem1903thermodynamics}{Duhem, P. M. M., \textit{Thermodynamics and Chemistry: A Non-Mathematical Treatise for Chemists and Students of Chemistry}, 1st ed.; J. Wiley \& sons: New York, 1903, p. 67.}

\bibitem{Meany1978}{Meany, J. E., ``Process Reversibility, Work, and Entropy.'' J. Chem. Educ. \textbf{55} (4), 238 (1978).}

\bibitem{VanNess1983}{Van Ness, H. C., \textit{Understanding Thermodynamics}; Dover Publications, Inc.: New York, 1983, p. 21.}

\bibitem{Thomsen1960a}{Thomsen, J. S., ``Distinction between Quasi-Static Processes and Reversibility.'' Am. J. Phys. \textbf{28} (6), 564--565 (1960).}

\bibitem{Thomsen1960}{Thomsen, J. S., ``Thermodynamics of an Irreversible Quasi-Static Process.'' Am. J. Phys. \textbf{28} (2), 119--122 (1960).}

\bibitem{Samiullah2007}{Samiullah, M., ``What Is a Reversible Process?'' Am. J. Phys. \textbf{75} (7), 608--609 (2007).}

\bibitem{Atkins2010}{Atkins, P.; De Paula, J., \textit{Physical Chemistry}; 9ts ed., W.H. Freeman and Company: New York, 2010; p. 51.}

\bibitem{Walker2013}{Walker, J.; Halliday, D.; Resnick, R., \textit{Fundamentals of Physics Extended}, 10 th.; John Wiley \& Sons, Inc., 2013.}

\bibitem{Fermi1956}{Fermi, E., \textit{Thermodynamics}; Dover Publications, Inc.: New York, 1956.}

\bibitem{Smith1996}{Smith, J. M.; Van Ness, H. C.; Abbott, M., \textit{Introduction to Chemical Engieneering Thermodynamics}, 6th ed.; McGraw-Hill, Inc., 1996.}

\bibitem{Rechel1947}{Rechel, E. E., ``The Reversible Process in Thermodynamics.'' J. Chem. Educ. \textbf{24} (6), 298--301 (1947).}

\bibitem{Landau1969}{Landau, L. D.; Lifshitz, E. M. \textit{Course of Theoretical Physics, Vol.\ 5: Statistical Physics.}; 2nd revised and enlarged ed., Pergamon Press: Oxford, 1969; p. 32.}

\bibitem{Zemansky1996}{Zemansky, M. W.; Dittman, R. H., \textit{Heat and Thermodynamics}, 7th ed.; McGraw-Hill, Inc.: New York, 1996.}

\bibitem{Callen1985}{Callen, H. B., \textbf{Thermodynamics and an Introduction to Thermostatistics}, 2nd ed.; John Wiley \& Sons.: New York, 1985.}

\bibitem{Thomsen1996}{Thomsen, J. S., ``The Reversible Process: A Zero-Entropy-Production Limit.'' Am. J. Phys. \textbf{64} (5), 580--583 (1996).}

\bibitem{Gislason2002}{Gislason, E. A.; Craig, N. C., ``First Law of Thermodynamics; Irreversible and Reversible Processes.'' J. Chem. Educ. \textbf{79} (2), 193--200 (2002).}

\bibitem{Ritacco2014}{Ritacco, H. A.; Fortunatti, J. C.; Devoto, W.; Fern\'andez-Miconi, E.; Dominguez, C.; Sanchez, M. D. ``Thermodynamics Fundamental Equation of a “Non-Ideal” Rubber Band from Experiments.'' J. Chem. Educ. \textbf{91} (12), 2195--2199 (2014).}

\bibitem{Ferrari2010}{Ferrari, C.; Gruber, C. ``Friction Force: From Mechanics to Thermodynamics.'' Eur. J. Phys. \textbf{31} (5), 1159--1175 (2010).}

\bibitem{Sherwood1984}{Sherwood, B. A. ``Work and Heat Transfer in the Presence of Sliding Friction.'' Am. J. Phys. \textbf{52} (11), 1001--1007 (1984).}

\bibitem{Besson2001}{Besson, U., ``Work and Energy in the Presence of Friction: The Need for a Mesoscopic Analysis.'' Eur. J. Phys. \textbf{22} (6), 613--622 (2001).}


\end{thebibliography}
\end{document}